\documentclass[%
 rsi,
 amsmath,amssymb,
 reprint,%
 groupedaddress,%
 superscriptaddress,
]{revtex4-1}

\usepackage{graphicx}
\usepackage{dcolumn}
\usepackage{bm}
\usepackage{algorithmic}
\usepackage{textcomp}
\usepackage{booktabs}
\usepackage{amsfonts,dsfont}
\usepackage{xcolor} 
\usepackage[caption=false,font=footnotesize]{subfig}
\usepackage[T1]{fontenc}
\usepackage{mathptmx}
\begin{document}

\title[]{Software-defined optoacoustic tomography}

%

\author{R. M. Insabella}
\affiliation{Universidad de Buenos Aires, Facultad de Ingenier\'ia, Paseo Col\'on 850, C1063ACV, Buenos Aires, Argentina.}%
\author{M. G. Gonz\'alez}%
\email{The author to whom correspondence may be addressed: mggonza@fi.uba.ar}%
\affiliation{Universidad de Buenos Aires, Facultad de Ingenier\'ia, Paseo Col\'on 850, C1063ACV, Buenos Aires, Argentina.}%
\affiliation{Consejo Nacional de Investigaciones Cient\'ificas y T\'ecnicas, (CONICET), Godoy Cruz 2290, C1425FQB, Buenos Aires, Argentina.}%
\author{L. M. Riob\'o}%
\affiliation{Universidad de Buenos Aires, Facultad de Ingenier\'ia, Paseo Col\'on 850, C1063ACV, Buenos Aires, Argentina.}%
\author{K. Hass}%
\affiliation{Universidad de Buenos Aires, Facultad de Ingenier\'ia, Paseo Col\'on 850, C1063ACV, Buenos Aires, Argentina.}%
\author{F. E. Veiras}
\affiliation{Universidad de Buenos Aires, Facultad de Ingenier\'ia, Paseo Col\'on 850, C1063ACV, Buenos Aires, Argentina.}%
\affiliation{Consejo Nacional de Investigaciones Cient\'ificas y T\'ecnicas, (CONICET), Godoy Cruz 2290, C1425FQB, Buenos Aires, Argentina.}


\begin{abstract}
In this work we present the first application of software-defined optoelectronics (SDO) for bidimensional optoacoustic tomography (OAT). The SDO concept refers to optoelectronic systems where the functionality associated with the conditioning and processing of optical and electrical signals are digitally implemented and controlled by software. This paradigm takes advantage of the flexibility of software-defined hardware platforms to develop adaptive instrumentation systems. We implement an OAT system based on a heterodyne interferometer in a Mach-Zehnder configuration and a commercial software-defined radio platform (SDR). Here the SDR serves as a function generator and oscilloscope at the same time providing perfect carrier synchronization between its transmitter and receiver in a coherent baseband modulator scheme. Therefore, the carrier synchronization enables us to have a much better phase recovery. We study the performance of the OAT SDO system by means of different bidimensional phantoms and the analysis of the reconstructed images.
\end{abstract}

\maketitle

\section{Introduction}
\label{s:intro}

\begin{figure*} [t]
	\centering
	\includegraphics[height=7cm]{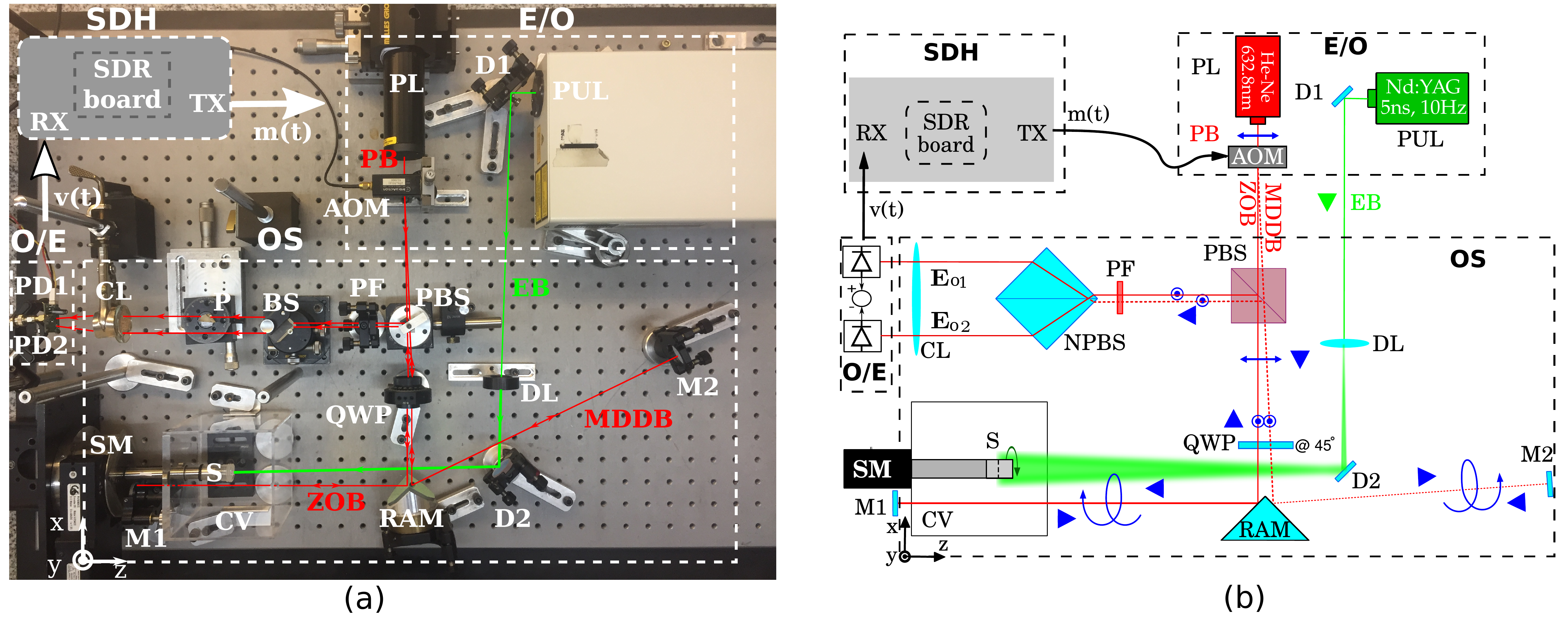}
	\caption{Software Defined Optical Interferometer (SDOI) setup. (a) Picture of the entire system. SDH: Software Defined Hardware block. $m(t)$: output electrical signal (TX). $v(t)$: input electrical signal (RX). E/O: Electrical-Optical block. PL: Probe Laser. PB: Probe Beam. AOM: Acousto-Optic Modulator. PUL: Pump Laser. DL: Diverging Lens. D: dichroic mirrors. EB: Excitation Beam. OS: Optical Setup block. PBS: Polarized Beam Splitter. QWP: Quarter Waveplate. RAM: Right Angle Mirror. M: mirrors. CV: Control Volume (cuvette). S: Sample. SM: Stepper Motor. PF: Passband Filter. BS: non-polarized Beam Splitter. CL: Converging Lens. P: prism. ZOB: Zero Order Beam. MDDB: Modulation Dependent Diffracted Beam. O/E: Optical-Electrical block. PD: Photodiodes. (b) SDOI block diagram.}
	\label{fig:setup}
\end{figure*}

The optoacoustic (OA) technique is the generation of acoustic waves due to thermoelastic expansion caused by absorption of short optical pulses. When the OA technique is used to perform a tomography system (OAT), the pressure profiles generated by the optical excitation are captured with ultrasonic sensors that surround the area of interest. A key element of any OA configuration is the detection system for sensing the acoustic waves. In OAT, ultrasound detectors can be classified in two categories, namely piezoelectric and optical detectors \cite{lutzwieler2013}. The former type is the most commonly used and it is based on polymeric (broadband), ceramic (resonant), or semiconductors (in the case of capacitive sensors) materials. Nowadays, especially for large-scale manufacturing, the piezoelectric technology has lower cost than the optical sensors. Moreover, capacitive micro-machined transducers are enhancing the development high-density detector arrays but they are much more expensive to develop in optical technology \cite{wissmeyer2018}. Nevertheless, optical ultrasound detection owns several advantages in biomedical applications, such as non-contact and remote inspection, sensitivity not dependent on the sensing area, optical transparency, material flexibility, and immunity to electromagnetic interference \cite{dong2017}.

One of the most used optical methods applied for detecting OA signals is interferometry \cite{wissmeyer2018}. Interferometric methods detect changes in optical interference patterns induced by ultrasound. Depending on the implemented interferometric configuration, perturbations in the interference pattern may be generated by changes in the mean free path, the optical phase or the optical wavelength. The resulting changes in intensity or frequency at the interferometer output are detected commonly by a photodiode. For example, in Michelson and Mach-Zehnder interferometers, the probe laser beam is split into two optical paths, one of which is perturbed by the ultrasound wave and the other serves as a reference. The two beams are combined at the interferometer output and their interference is measured. In these schemes, an acoustic coupling medium must be used when ultrasound interacts with the beam path. Generally, the sensitivity of an interferometric ultrasound sensor is determined by the efficiency with which acoustic perturbations are converted into changes in light characteristics in the optical system and by the responsivity of the photodiode for detecting those changes. Two-beam interferometers are usually implemented in fiber-based \cite{bauer2017} or free-beam \cite{paltauf2007photoacoustic,nuster2010} configurations, and most of them rely on continuous-wave (CW) lasers. The interferometers based on CW lasers have two main disadvantages in OAT \cite{wissmeyer2018}: i) it is very difficult and expensive to detect OA signals at different places at the same time, and ii) these systems are sensitive to temperature drifts and vibrations, such as motion of large samples during in vivo imaging. An improvement on the last issue can be achieve using frequency-modulation techniques, such as heterodyne detection \cite{park2014}.

In this work, we present the first application of the software-defined optoelectronics (SDO) concept \cite{riobo2018} for bidimensional OAT. The SDO concept refers to optoelectronic systems in which most of the functionality associated with signal conditioning and processing is digitally implemented and controlled by software in real time. This includes the optical signal modulation and demodulation, and coding and decoding, with minimum hardware modifications \cite{riobo2018}. Usually OAT setups are based in a fixed hardware implementation and most embedded systems used in instrumentation have a dedicated function for which they are optimized. That scheme provides little flexibility when prototyping and the experiments tend to be very hardware dependent. In contrast, SDO architecture makes use of a general purpose radio frequency software defined hardware platform which can be modified in real time by means of software. The difference between SDO and conventional (mature) testing systems is that when using a dedicated hardware such as an oscilloscope, or any other embedded system such as FPGA, there are hardware constraints associated with the instrumentation system which limit the functionality of most experiments. Our OAT-SDO system includes a software-defined balanced-path heterodyne interferometer as the optical sensor based on a software defined radio (SDR) platform \cite{riobo18_2} to perform the optical modulation and demodulation of the interferometric signals and the decoding of the ultrasound signal. Here the SDR serves as a function generator and oscilloscope at the same time providing perfect carrier synchronization between its transmitter and receiver in a coherent baseband modulator scheme. The carrier synchronization enables us to have a much better phase recovery in our heterodyne interferometer.

The paper is organized as follows. In Sect.~\ref{s:SDOI} we present the implemented software defined interferometer (SDOI) system based on a commercial SDR and a Mach-Zehnder configuration interferometer. In Sect.~\ref{s:OI} we describe the application of the SDOI system to perform 2-D OA images. We provide a description of the methods and materials employed and we also present the results obtained from several measurements made over different phantoms. Finally, in Sect.~\ref{s:conclu}, we present the conclusions and give a brief overview regarding possible future developments based on this technology. 

\begin{figure*} [t]
	\centering
	\includegraphics[height=7cm]{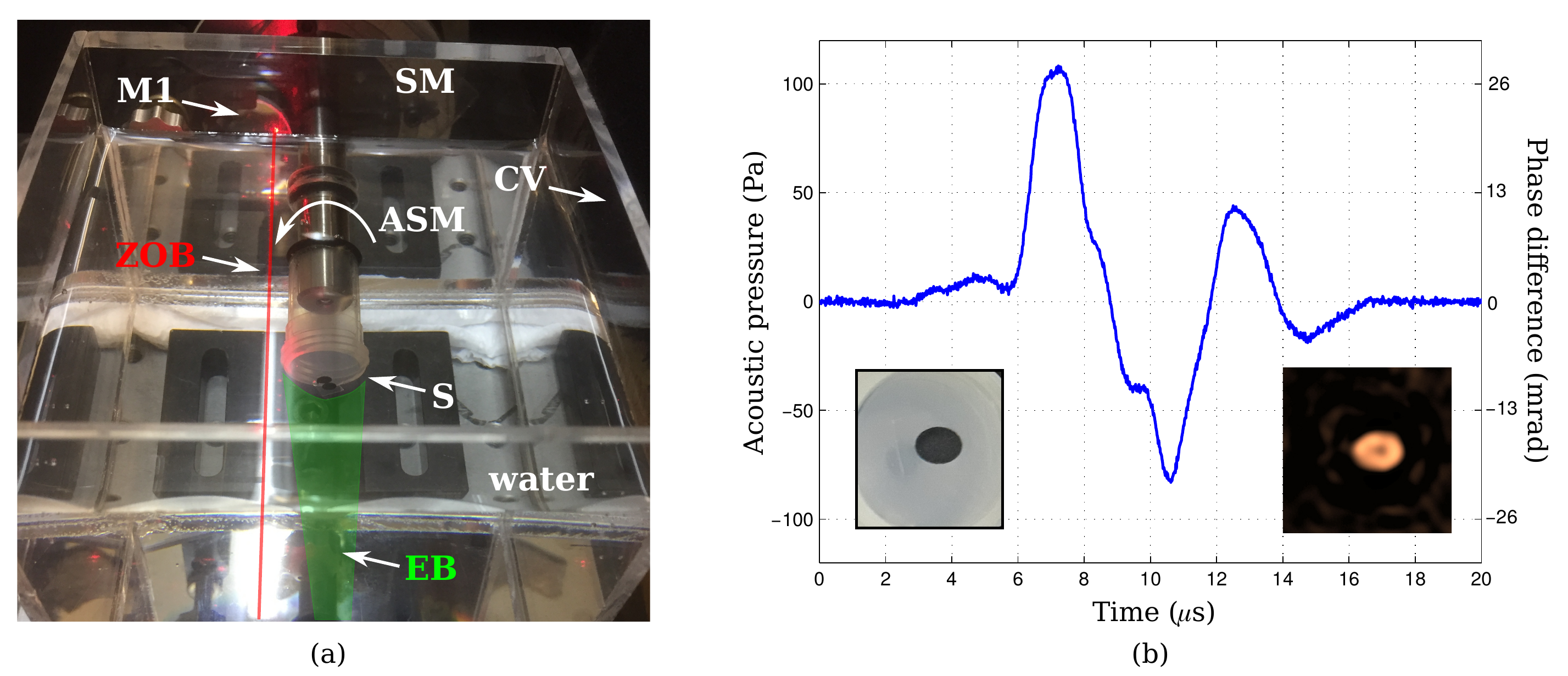}
	\caption{(a) Picture of the control volume (CV) with one of the samples used in this work showing how looks the samples used in this work (ink pattern printed in a transparent film attached to an agarose gel cylinder). ASM: Axis of the Stepper Motor. (b) OA signal generated from a black elliptical disk of 3$\times$4 mm irradiated by a single laser pulse of 8 mJ/cm$^2$ and captured by the SDOI at a certain angle. Insets: ink pattern in agarose gel (left) and reconstructed image (right).}
	\label{fig:example-signal}
\end{figure*}

\section{Software Defined Optical Interferometer}
\label{s:SDOI}

The optical detector used in our OAT-SDO system is a software defined optical interferometer (SDOI), whose experimental setup is shown in Fig. \ref{fig:setup}. In its design we considered four fundamental elements: the optical system (OS), the electrical-optical block (E/O), the optoelectronic block (O/E) and the software defined hardware (SDH). The latter controls both electrical-optical and optoelectronic blocks in order to define the functionality of the optical system. The electrical-optical block consist of a horizontally polarized HeNe laser source (632.8 nm, 10 mW) and an acousto-optical modulator (AOM) connected to the transmitter of a SDR platform. The SDR drives the acousto-optical modulator with a carrier signal $m(t)$ of frequency $f_0$, producing two beams with different optical frequencies: the zero order beam (ZOB) and the modulation-dependent diffracted beam (MDDB). These beams enter to the optical system. The electrical-optical block also provides the optical system with pulsed laser radiation (EB) to excite the sample (S) in the control volume (CV). In the optical system, the optical beams from the electrical-optical block, ZOB and MDDB, are transmitted through a polarization beam splitter (PBS) and a quarter-wave plate (QWP) becoming circularly polarized. A right angle mirror (RAM) reflects the beams into different paths with similar lengths. The ZOB traverses the control volume (OA source) two times since it is reflected on the mirror (M1), whereas the MDDB reflects on a mirror (M2). Since both beams are reflected back towards the polarization beam splitter, traversing two times the quarter-wave plate, their polarization plane is rotated 90$^\text{o}$ becoming vertically polarized. Consequently, when the beams reach the polarization beam splitter, they are reflected and redirected to traverse a passband filter (PF) that is used to reject any other wavelength different from 633 nm $\pm$ 5 nm. Then, the beams are recombined by a single-element interferometer \cite{ferrari2007}, consisting of a 50/50 ratio beam splitter (BS), producing two interferometric signals in counterphase,

\begin{align}
I_1(\mathbf{r},t) = A(\mathbf{r},t)+B(\mathbf{r},t)\cos[\Delta\phi(\mathbf{r},t)] \\
I_2(\mathbf{r},t) = A(\mathbf{r},t)-B(\mathbf{r},t)\cos[\Delta\phi(\mathbf{r},t)]
\end{align}

\noindent where $A(\mathbf{r},t)$ is the background intensity, $B(\mathbf{r},t)$ is the fringe contrast and $\Delta\phi(\mathbf{r},t)$ is the phase difference between the beams. Any beam deviation is corrected by means of a rotatable prism (P) and the output beam-widths are finally adjusted by a converging lens (CL) (Fig. \ref{fig:setup} (a)) in order to match the optoelectronic block input. These interferometric signals are detected by the optoelectronic block, which consists of a custom-made balanced photodetector (PD1 and PD2) based on PIN photodiodes (model SFH2701) whose sensitive area is 0.36 mm$^2$. In order to measure $v(t)$, we employed a wideband transimpedance amplifier based on a JFET input operational amplifier \cite{riobo_17}. The whole front-end has a bandwidth of $120$ MHz and output impedance of 50 $\Omega$. Therefore, the input signal $v(t)$ from the optoelectronic block may be written as: 

\begin{equation}
v(t) \propto I_1(t)-I_2(t) = 2B(t)\cos[2\pi f_0 t + \Delta\phi(t)]
\label{OE_signal}
\end{equation}

\noindent where $f_0 =$ 75 MHz for this work.

The balanced photodetection takes advantage of the symmetry between the output optical signals from the optical system, allowing high common-mode rejection of the intensity fluctuations of the laser source and stray incoherent light. Moreover, since the optical path difference between the beams of the interferometer is carefully controlled to be nearly zero, the phase noise contribution from the laser source can be neglected. 

The OA wave within the control volume produces a non-homogeneous, time-dependent refractive index variation $\Delta n(\mathbf{r},t)$ in the propagating medium,
\begin{equation}
\Delta n(\mathbf{r},t) = \frac{dn}{dp} p(\mathbf{r},t)
\label{eqDn}
\end{equation}
\noindent where $p$ is the acoustic pressure generated by the OA effect and $dn/dp =$ 1.35 10$^{-10}$ Pa$^{-1}$ for water \cite{paltauf2007photoacoustic}. This spatio-temporal variation of the refractive index modifies the amplitude $B(t)$ of the interferograms (due to beam deflection, for example) and also produces a phase difference between the beams that traverse the control volume (due optical path variation). It is important to notice that the heterodyne technique applied in this work provides a way to decouple these effects over the detected signal. The phase difference depends on the integrated effects of the refractive index variations along the propagation paths follow by the ZOB and MDDB \cite{paltauf2007photoacoustic}. Therefore, 

\begin{equation}
\Delta \phi(t) = \frac{2\pi}{\lambda}L \Delta n(t)
\label{fase_zob_mddb}
\end{equation}

\noindent where $\lambda$ is the probe laser (PL) wavelength and $L$ the length of the propagation path in the control volume. 

The optoelectronic block is connected to the receiver port of the SDR platform (LimeSDR from LimeMicrosystems \cite{LimeSDR}). A generic laptop computer interfaces the SDR input and output data streams to the outside world through USB communication. It also manages the SDR software components and provides any additional control signals using the high-level programming language Octave. The LimeSDR is configured to perform modulation and demodulation of the signals $m(t)$ and $v(t)$. The receiver is configured by software to perform quadrature demodulation \cite{Stewart2015} in order to obtain the phase information. Here, the received signal is sampled (60 MS/s) and decomposed into its quadrature components

\begin{equation}
B(t)\{\cos[\Delta \phi(t)]+j\sin[\Delta \phi(t)]\}=v_I(t)+jv_Q(t)
\end{equation}

\noindent where the digitized instantaneous phase difference is retrieved by computing,

\begin{equation}
\Delta \phi(t) = \arctan\left[\frac{v_Q(t)}{v_I(t)}\right]
\label{demod_SC}
\end{equation} 

\section{Optoacoustic Imaging using SDOI}
\label{s:OI}

\begin{figure}
	\includegraphics[width=\columnwidth]{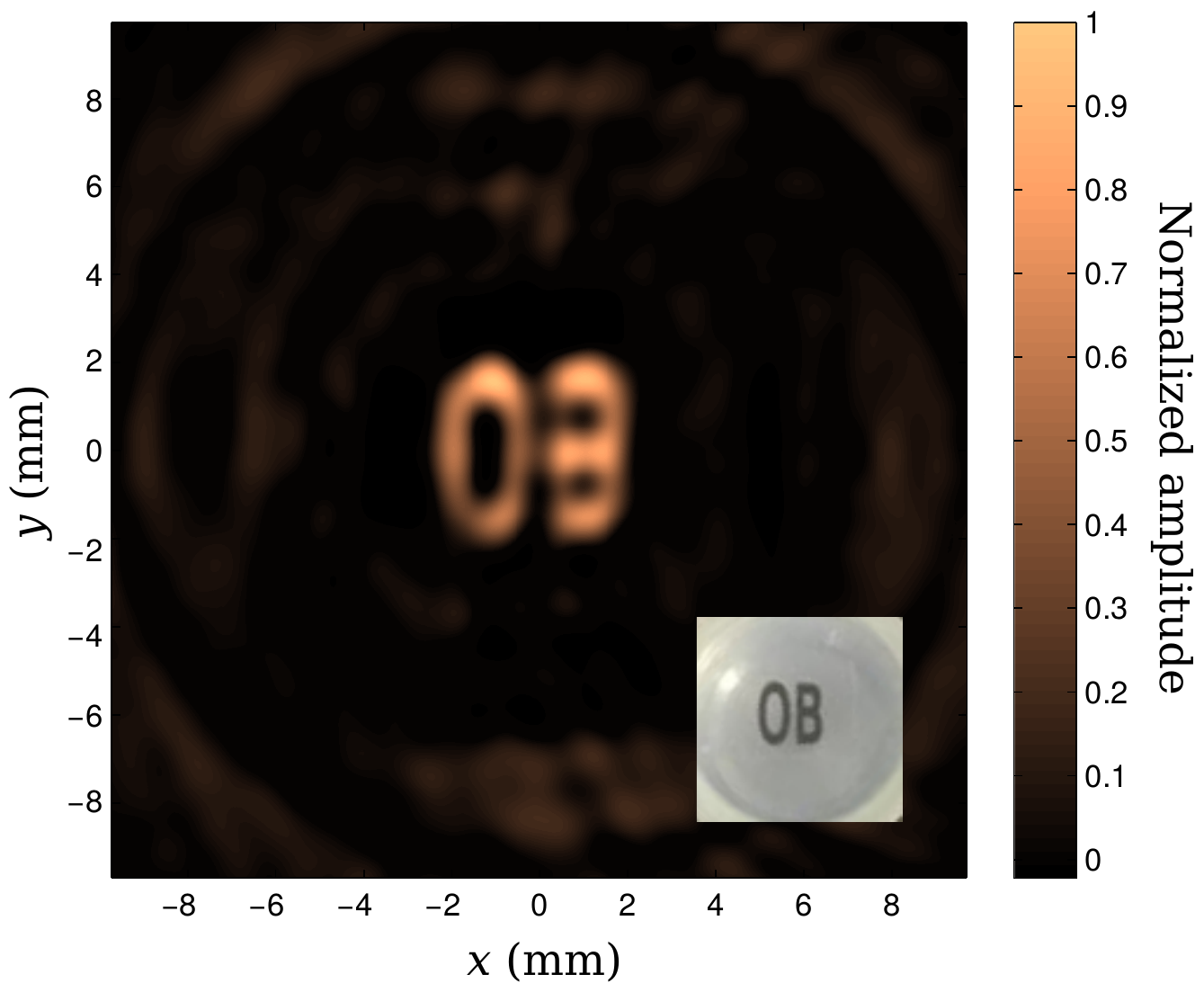} 
	\caption{Reconstructed image obtained with the OAT system under study. The phantom reads OB and its size is 4 mm x 4 mm. The sample (see inset) was irradiated with a laser fluence of 8 mJ/cm$^2$.}
	\label{fig:Imagen_OB}
\end{figure}

The implemented 2-D OAT system is based on the SDOI described in section. \ref{s:SDOI}.  The sample is immersed in a large square vessel (100 mm per side) filled with deionized water (control volume). As an excitation laser (PUL), we used a Nd:YAG laser (Continuum Minilite I, output wavelength: 532 nm) with a pulse duration of 5 ns, and a repetition rate of 10 Hz (see electrical-optical block in Fig. \ref{fig:setup}). In all the measurements we carried out in this work, the fluence is less than 10 mJ/cm$^2$. A diverging lens (DL) adapts the diameter of the laser beam to the size of the sample (14 mm of diameter), trying to achieve an homogeneous illumination in the system imaging plane. Phantoms are fixed to a rotatory stage (Newport PR50CC) and rotated 360$^{\circ}$ in 1$^\circ$ steps since full view data (i.e. 360$^\circ$) minimizes the effect of a limited view detection \cite{xu2004}. In order to improve the quality of the images, the OA signals are 4 times averaged for each angle. The distance between the axis of the stepper motor (ASM) and the ZOB is 13 mm (see Fig. \ref{fig:example-signal}(a)).

\begin{figure*} [t]
	\centering
	\includegraphics[height=7cm]{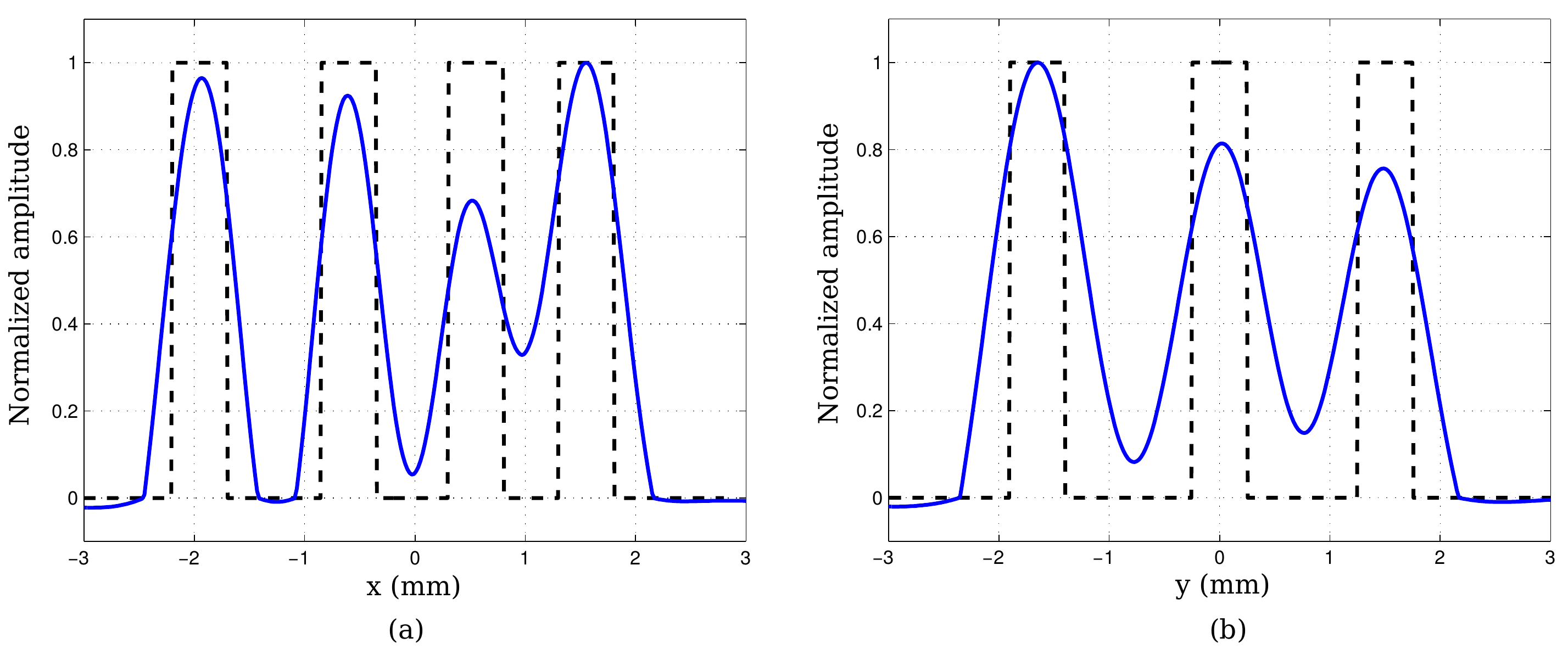}
	\caption{Intensity profile along a cross-section from Fig. \ref{fig:Imagen_OB}. Black dashed line: theoretical ink pattern. Blue solid line: reconstructed profile. (a) $y = -0.79$ mm.  (b)  $x$ = 1 mm. The theoretical ink pattern is not be perfectly rectangular but trapezoidal. However, these differences between trapezoidal and rectangular shapes are not perceptible in this scale.}
	\label{fig:ImageOBxyfijo}
\end{figure*}

The active detection area of the SDOI based sensor is approximately 0.9 mm x 200 mm, since the ZOB traverses the 100 mm length cuvette two times. These values allow us to achieve a homogeneous resolution in the scanned region \cite{paltauf2009}. Moreover, its effective signal bandwidth (BW) is proportional to the ratio between the square root of the sensitive area of the photodiodes and the speed of sound in the media (BW $\simeq$ 2 MHz) \cite{riobo2018}. This value was verified by analyzing the fourier transforms of the measurements.

The samples used to test our setup consist of ink patterns laser printed on transparent films embedded in agarose gel. The agarose gel is prepared with 2.5$\%$ (w/v) agarose in distilled water. First, a cylindrical base of the agarose gel with a diameter of 14 mm and a height of approximately 20 mm is prepared. Then, the object (ink pattern on transparent film) is placed in the middle of the cylinder and fixed with a few drops of the gel. Finally, another layer of gel with a thickness of $\sim$1 mm is formed on top of the sample object. We tested different ink patterns such as circular and elliptical disks, letters and numbers of various sizes. A picture of how the sample is mounted and the relative positions of the beams is shown in Fig. \ref{fig:example-signal}(a).

There are several techniques and procedures for the obtention of OA images. However, the approach that has had the best experimental results is the back-projection (BP) technique which is a time domain algorithm very simple to implement \cite{rosenthal2013}. For this reason, we decide to use this approach. There are many implementations of this algorithm. Since our optical detector is an integrating line sensor, we implemented a BP algorithm following the procedure detailed in ref. \cite{burgholzer2007}. In order to reconstruct 2-D images, signals are measured at each angular position of the sample. Given the integrating property of the line detectors, these images are projections of the initial pressure distribution in the sample into a plane $xy$ (see Fig. \ref{fig:example-signal} (a)). This distribution is directly proportional to the absorbed energy density. 
Here the recorded pressure signals are back projected onto cylinders with the line sensor associated to ZOB in their center. Before back projection, the captured signals are de-noised by applying a low-pass filter with a cut-off frequency of 2 MHz (i.e., the same value of the detection system bandwidth, BW). In order to convert these time signals into distance from the line sensor, the BP algorithm requires the exact value of the sound speed. Due to the temperature dependence of the sound speed it is necessary to determine the exact temperature of the water bath in each experiment. Therefore, the water temperature is measured with a calibrated thermocouple. We studied the general performance of our system by means of different phantoms (ink patterns of letters, numbers and geometrical shapes). 

The implemented optical heterodyne interferometer exploits the SDR capacities serving as a function generator and oscilloscope at the same time providing perfect carrier synchronization between its transmitter and receiver. The carrier synchronization allows a better phase recovery and measurements with large signal to noise ratio (SNR). An example of this feature is presented in Fig. \ref{fig:example-signal} (b) where the OA signal generated from a black elliptical disk (3x4 mm) irradiated by a single pulse laser of 8 mJ/cm$^2$ and captured by the SDOI at a certain angle is plotted. In the same figure it can be appreciated a picture of the sample and the reconstructed image. As it can be seen, the left axis is graduated in mrads and it is possible to appreciate the performance of the system regarding phase measurement. It is important to note that we achieved a high speed very precise phase measurement. Here, the OA pulsed signals produce peaks of tens of mrads that can be measured with an small uncertainty of less than 1 mrad (relative uncertainty is less than 1$\%$).

In order to obtain the sensitivity of the detection system, we used a calibrated piezoelectric sensor \cite{abadi2017,abadi2018}. We recorded the OA signal generated by a sample using both detectors (SDOI and piezoelectric). Comparing those signals, we determined a sensitivity value of 0.27 mrad/Pa. Then, we estimate the SDOI measurement system noise floor through the histogram of the detected phase difference $\Delta \phi$ without any phantom (i.e. only the transparent agarose gel is on the control volume). The histogram corresponds to a Gaussian-type distribution of zero mean. From its fitted standard deviation, the estimated noise floor for a single pulse laser was 0.55 mrad. From these values (sensitivity and noise floor), we obtained a noise equivalent pressure (NEP) over the system bandwidth (2 MHz) of $\sim$2 Pa.

This is a result of the combination of heterodyne interferometry with a radio receiver (SDR) that is capable of demodulate small phase modulations. The system itself measures high-speed pressure waves by means of refractive index variations. It is also important to note that the refractive index variation associated to the peaks of the OA signals are about 1$\times 10^{-8}$ RIU, making this system very attractive for evaluation of material properties. 

To estimate the spatial resolution of the device, we analyzed the OA images for different phantoms (letters, numbers, shapes, etc.) embedded in agarose gel. An example of a reconstructed image is presented in Fig. \ref{fig:Imagen_OB}. This image is obtained from a phantom based on two capital letters (OB).

The reconstructed image clearly allow us to properly identify both characters. We compare the full width at half maximum (FWHM) of the intensity profiles between the theoretically printed ink pattern and the reconstructed image. An example of this procedure is shown in Fig. \ref{fig:ImageOBxyfijo} (a) and (b) where the intensity profile from Fig. \ref{fig:Imagen_OB} is analyzed along a horizontal and a vertical line, respectively.

From Fig. \ref{fig:Imagen_OB}, we obtained several values of FWHM (4 for $x$ and 3 for $y$). We repeated this procedure on 10 phantoms obtaining approximately 40 values of FWHM for each axis. These values are approximately equal in $x$ and $y$ and lies between 700 and 800 $\mu$m. Considering the finite size of the ink patterns, this can be regarded as an upper limit of the resolution \cite{paltauf2017}. It is interesting to compare this value with the ideal maximal achievable resolution, $R_{bw}$. Using the values of the sound speed measured in this work ($v_s=$ 1480 m/s) and the bandwidth of the SDOI ($BW=$ 2 MHz), $R_{bw} \approx 0.8 v_s/BW=$ 590 $\mu$m  \cite{xu2003}. It is important to note that this expression assumes an idealized scenario, i.e. full view detection, point detector, continuous spatial sampling and constant sound speed that can not be fully accomplished in these experiments and therefore it gives rise to this difference.

\section{Conclusions}
\label{s:conclu}

In this article we demonstrate that it is possible to apply the SDO concept to obtain OA images. We implement a 2-D OAT system based on a heterodyne interferometer in a Mach-Zehnder configuration and a commercial SDR platform. Moreover, it is important to note that, within our knowledge, this is the first OAT system based on SDR technology. Future advances in OAT SDO systems will probably include more control of the excitation beams, for example by means of intensity-modulated cw laser diodes \cite{maslov2008photoacoustic}.

We study the system performance by means of measurements of different bidimensional phantoms. From the analysis of the results we obtain that this scheme has a very good sensitivity (NEP density = 1.41 mPa/Hz$^{1/2}$) and an adequate resolution of 750 $\pm$ 50 $\mu$m. Moreover, the interferometric performance regarding phase measurement make the system also very attractive for the analysis of material properties since it can measure refractive index variations up to the eight figure. Peaks ranging a few mrads are easily detected over a noise floor that does not exceed 0.55 mrad. Such sensitivity is obtained by means of a low cost general purpose radio frequency software defined hardware platform which is comparable to those obtained by similar methods \cite{paltauf2007photoacoustic}. In terms of NEP density, our system undercuts previously reported values for fiber based optical detection of ultrasound (such as 7.6 mPa/Hz$^{1/2}$ in \cite{bauer2017}).

The system shown in this work has not only the inherent advantages of optical sensors such as noncontact and remote inspection, optical transparency, and immunity to electromagnetic interference, but also the flexibility of the SDR technology which allows to implement adaptive optoelectronics systems which may be configured in real time. By means of example, changing the carrier frequency of the SDR transmitter implies a change in the deflection angle at the output of the optoacoustic modulator changing the spatial position of the probe beam. Another example is that it can be set up the size of the memory bus or the cut-off frequency of the low pass filter at the receiver while performing an experiment. Particularly, in this article, the SDR platform serves as a function generator and oscilloscope at the same time providing perfect carrier synchronization between its transmitter and receiver in a coherent baseband modulator scheme and advantages in both flexibility and cost.

One of the main challenges is to improve the spatial resolution of the OAT systems. We believe that similar setup using fiber optics instead of free space propagating beams will improve the spatial resolution of the OAT systems. Moreover, the advantages of the SDR developing a multicarrier heterodyne interferometry \cite{riobo2018} have shown promising results that could be exploited in fiber optic interferometers \cite{riobo2019noise}. It improves the noise floor and enables the discrimination between external mechanical disturbances and electrical noise.  Moreover, the multicarrier modulation introduces such a space and frequency diversity that makes  it possible to detect an OA signal in different places for a single laser pulse, thus obtaining OA images in almost real time and with a lower cost.

\subsection* {Acknowledgments}
This work was supported by the University of Buenos Aires (UBACyT grants: 20020160100052BA, 20020160100042BA, 20020170200232BA) and the ANPCyT (PICT grant 2016-2204).

\subsection* {Data Availability Statement}
The data that support the findings of this study are available from the corresponding author upon reasonable request.



\bibliography{Gonzalez_etal_references}


\end{document}